\documentclass[twocolumn]{aastex62}
\usepackage{graphicx,ifthen,url,float,color,ulem,amsmath,ragged2e}

\def\ltsima{$\; \buildrel < \over \sim \;$}
\def\simlt{\lower.5ex\hbox{\ltsima}}
\def\gtsima{$\; \buildrel > \over \sim \;$}
\def\simgt{\lower.5ex\hbox{\gtsima}}
\newcommand {\uJy}{$\mu$Jy}
\newcommand {\um}{$\mu$m}

\newcommand{\lsun}{{\rm\,L$_\odot$}}

\newcommand{\lpeak}{$\lambda_{\rm peak}$}
\newcommand{\lir}{L$_{\rm IR}$}

\newcommand\mmpz{{\sc MMpz}}

\shorttitle{\mmpz: FIR photometric redshifts}
\shortauthors{Casey}

\begin{document}

\title{\sc Far-Infrared Photometric Redshifts: A New Approach to a Highly Uncertain Enterprise}

\correspondingauthor{Caitlin M. Casey}
\email{cmcasey@utexas.edu}

\author[0000-0002-0930-6466]{Caitlin M. Casey}
\affil{Department of Astronomy, The University of Texas at Austin, 2515 Speedway Blvd Stop C1400, Austin, TX 78712, USA}

\begin{abstract}
I present a new approach at deriving far-infrared photometric
redshifts for galaxies based on their reprocessed emission from dust
at rest-frame far-infrared through millimeter wavelengths.
Far-infrared photometric redshifts (``FIR-$z$'') have been used over
the past decade to derive redshift constraints for highly obscured
galaxies that lack photometry at other wavelengths like the
optical/near-infrared.  Most literature FIR-$z$ fits are performed
through $\chi^2$ minimization to a single galaxy's far-infrared
template spectral energy distribution (SED).  The use of a single
galaxy template, or modest set of templates, can lead to an
artificially low uncertainty estimate on FIR-$z$'s because real
galaxies display a wide range in intrinsic dust SEDs.  I use the
observed distribution of galaxy SEDs (for well-constrained samples
across $0<z<5$) to motivate a new far-infrared through millimeter
photometric redshift technique called \mmpz.  The \mmpz\ algorithm
asserts that galaxies are most likely drawn from the empirically
observed relationship between rest-frame peak wavelength, \lpeak, and
total IR luminosity, \lir; the derived photometric redshift accounts
for the measurement uncertainties and intrinsic variation in SEDs at
the inferred \lir, as well as heating from the CMB at $z\simgt5$.  The
\mmpz\ algorithm has a precision of $\sigma_{\Delta
  z/(1+z)}\approx0.3-0.4$, similar to single-template fits, while
providing a more accurate estimate of the FIR-$z$ uncertainty with
reduced chi-squared of order $\mathcal{O}(\chi^2_{\nu})=1$, compared
to alternative far-infrared photometric redshift techniques (with
$\mathcal{O}(\chi^2_\nu)\approx10-10^3$).
\end{abstract}


\section{Introduction} \label{sec:intro}

Galaxies' far-infrared spectral energy distributions are notoriously
undersampled, whether or not the galaxies sit at $z=0$ or $z=7$.  Data
is sparse in this far-infrared through millimeter wavelength range
(herein referred to as FIR/mm, at rest-frame $\sim$20\um--3\,mm)
because Earth's atmosphere is largely opaque at these wavelengths;
thus, we rely on insight from the limited far-infrared space-based
missions (e.g. {\it Spitzer}, {\it IRAS}, {\it ISO} and
the {\it Herschel Space Observatory}), or the limited view we can
achieve through the handful of atmospheric windows we can peer through
from the ground (e.g. with the James Clerk Maxwell
  Telescope, JCMT, the NOrthern Extended Millimeter Array, NOEMA, and
  the Atacama Large Millimeter and submillimeter Array, ALMA).

The FIR/mm wavelength regime is sensitive to reprocessed stellar
emission, absorbed and re-radiated by dust in the interstellar medium
(ISM), and is known to host a rich suite of spectral features useful
for diagnostics of galaxies' gas, metal and dust content.  The most
prominent characteristic of this regime is the superposition of
modified blackbodies (of different temperatures and luminosities),
originating from diffuse dust in the ISM.  Most of the dust mass is
relatively cold ($\approx$20-60\,K), resulting in a peak of the SED
around rest-frame \lpeak\,=\,100$\pm$50\um.  In the local Universe,
there is a noted variation in SEDs' luminosity-weighted dust
temperature $T_{\rm d}$, a quantity that scales inversely with the
observed rest-frame peak wavelength $\lambda_{\rm peak}$: from the
$\sim$18\,K dust in the Milky Way and similar L$_\star$ galaxies to
the $\sim$80\,K dust in the dust-enshrouded Arp\,220 or $\sim$60\,K
dust in the starburst galaxy M82.  That same variance is seen at
higher redshifts, with a general trend between \lpeak\ and \lir, such
that intrinsically more luminous galaxies are also hotter
\citep[e.g.][]{chapman04a,casey12a,kirkpatrick12a,casey18a}.  However,
the relationship between \lpeak\ and \lir\ has significant scatter,
attributable to variable dust geometries.

Despite the measured variance in galaxies' dust SEDs, it is common for
0$^{th}$ order approximations for dust emission to take hold when
there is little to no data available to analyze.  For example, this is
done especially when a source's redshift is unconstrained and there is
little to no other data (like a spectrum or photometric data in the
radio/optical/near-infrared) available to constrain the redshift.  In this
case, the limited information available at FIR/mm wavelengths is used
directly to place constraints on the redshift, a technique called
far-infrared photometric redshift fitting (hereafter FIR-$z$ fitting).
While this FIR-$z$ fitting technique is simple in its application,
systematic offsets are especially problematic if the intrinsic dust
SED of the fitted galaxy is significantly different than the
template used to derive the FIR-$z$.  FIR-$z$ fits also tend to have underestimated
uncertainties due to a lack of accounting for the underlying variance
in dust SEDs.

In this paper, I introduce a new approach at fitting far-infrared
through millimeter photometric redshifts, using the empirical
relationship between \lir\ and \lpeak, as well as photometric
uncertainty.  I call this fitting technique ``\mmpz,'' shorthand for
millimeter photometric redshift\footnote{In this paper far-infrared
  and millimeter are used somewhat interchangeably to refer broadly to
  a galaxy's dust SED extending from $\sim$5\,\um--3\,mm.}.  
The code is made available for public use
\footnote{www.as.utexas.edu/$\sim$cmcasey/mmpz.html} including example
use cases for well-known dusty star-forming galaxies. I compare this
photometric redshift method with the use of single galaxy template
FIR-$z$ fits, and related alternate FIR-$z$ techniques from the
literature, with particular focus on their predicted uncertainties and
accuracy.  A brief history of FIR-$z$ fitting in the literature is
given in \S~\ref{sec:techniques}, while \S~\ref{sec:mmpz} describes
the \mmpz\ approach to FIR-$z$ fitting.  Quantitative tests of both
techniques using mock and limited real data are
described in \S~\ref{sec:tests}, and conclusions are described in
\S~\ref{sec:conclusions}.

\begin{figure}
  \includegraphics[width=0.99\columnwidth]{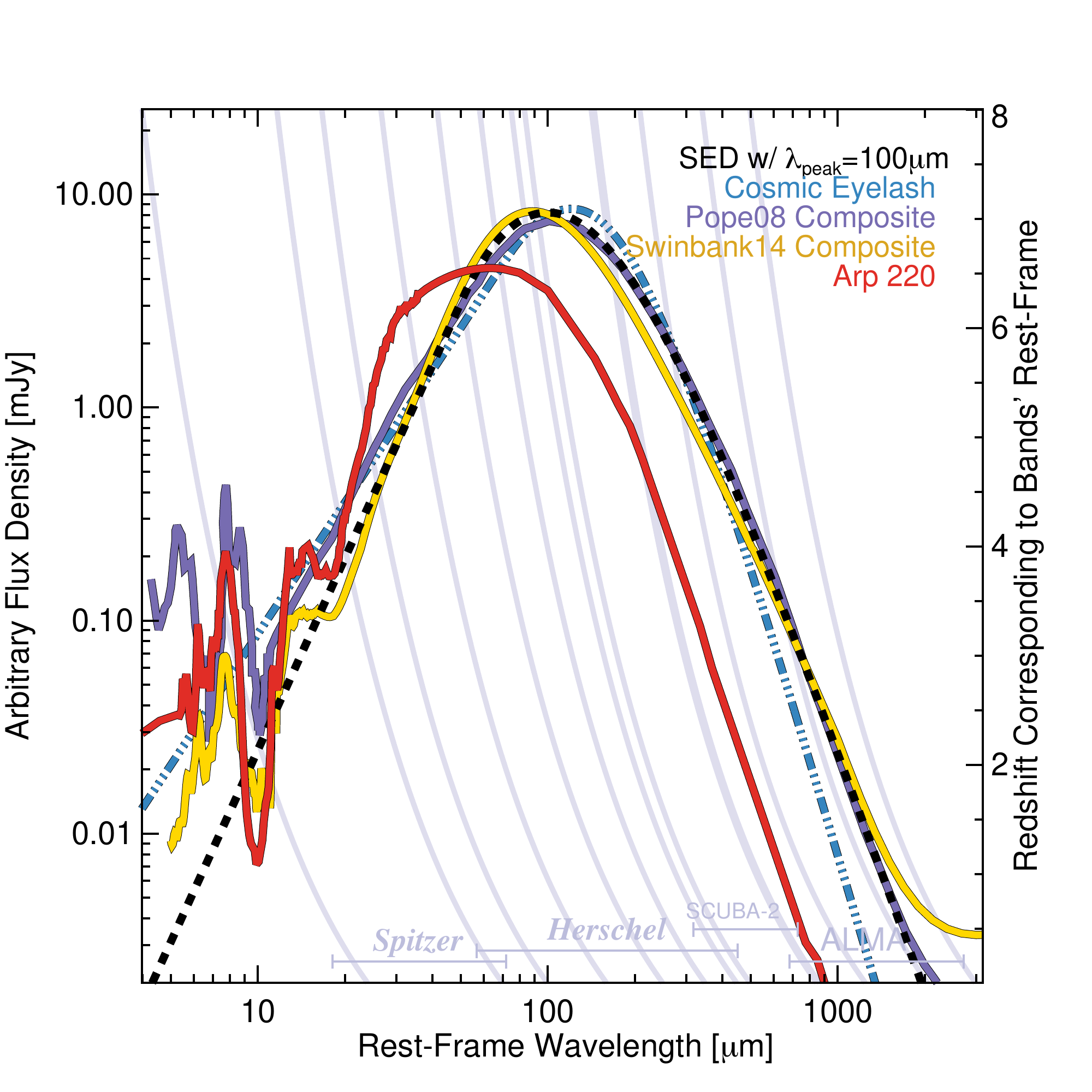}
  \caption{Example SEDs that are commonly used in the literature as
    representative of galaxies' long wavelength SEDs, where the
    integral under the curve corresponds to the obscured
    star-formation rate via \lir\,$\propto$\,SFR.  All SEDs are
    normalized to \lir=$10^{12}$\,\lsun.  Arp\,220 is in red
    \citep{klaas97a,rigopoulou96a}, the Cosmic Eyelash in dot-dashed blue
    \citep{swinbank10a}, the \citet{pope08a} SMG composite in purple,
    the \citet{swinbank14a} composite in gold, and a generic
    modified black body plus mid-infrared powerlaw that peaks at
    rest-frame $\lambda_{\rm peak}=100$\um in dashed black.  The right
    axis and light gray lines show the wavelengths of common FIR/mm
    continuum measurements and their corresponding rest-frame
    wavelengths at different redshifts.}
  \label{fig:sampleseds}
\end{figure}

\section{Literature Techniques}\label{sec:techniques}

Several works have used FIR photometric redshift fitting techniques to
assess the redshifts of sources otherwise unconstrained through other
data, dating back to the first observations of the first distant DSFGs
in the late 1990's.  This section describes some of those methods
broadly. Later in \S~\ref{sec:tests}, we compare between some of these
techniques --- in particular those that are not reliant on any
photometric measurements outside of the FIR/mm regime (including radio
wavelengths) --- and the new approach outlined in \S~\ref{sec:mmpz}.

 Some of the first studies recognizing the redshift
  evolution of submillimeter colors, thus the potential to use such
  measurements to constrain redshift, came in the very early days of
  bolometer observations at the JCMT. \citet{hughes97a} modeled tracks
  of cold dust emission in 400\um/800\um\ color (then observed with
  the single-element bolometer UKT14) demonstrating strong redshift
  evolution beyond $z\sim1$.
\citet{barger00a} combined the first 850\um\ datasets from
      {\sc Scuba} with existing deep \uJy-radio imaging datasets from
      the VLA to derive a photometric redshifts based on an assumption
      of the FIR-radio correlation.  Without detailed FIR SED
      constraints beyond the single 850\um\ {\sc Scuba} point, they
      used an Arp\,220 template
      \citep{klaas97a,rigopoulou96a,condon91a} to tether the data,
      recognizing that such a template could be used to predict
      redshift from the 353\,GHz-to-1.4\,GHz flux density ratio.  This
      technique relied on the assumption that the FIR-radio
      correlation \citep{helou85a} holds to high-redshift
      \citep[c.f.][]{delhaize17a}. The range of plausible redshifts
      for each source was then determined using a $\chi^2$ maximum
      likelihood technique.
\citet{aravena10b} use a similar FIR-radio-$z$ fitting method to for
DSFGs discovered in the COSMOS field with unconstrained redshifts
similarly focusing on the combination of the millimeter and radio
continuum and use of Arp\,220 as a prototypical template for the
higher redshift DSFGs being fit.

Both of these approaches were practical given the available data at
the time.  In the 2000's, DSFGs usually were only identified at one
wavelength in the FIR/mm regime, and the use of radio counterpart
identification was commonly used to astrometrically identify the most
likely multiwavelength counterparts.  The use of Arp\,220 in these
techniques was also sensible as it is a galaxy in the local Universe
that sits in the same luminosity class of many of the analyzed
high-$z$ DSFGs.  The combination of a single FIR/mm photometric
constraint with one radio photometric constraint can be an effective
tool to constrain redshift due to the different $K$-corrections
applied to either wavelength regime.

A more complex approach is outlined in \citet{aretxaga03a}, who use a
set of 20 SEDs with different rest-frame peak wavelengths to generate
a distribution in colors for galaxies as a function of redshift, drawn
from an assumed integrated IR luminosity function (IRLF).  The
probability distribution in redshift for any given source of measured
850\um-to-1.4\,GHz color (or 1.2\,mm-to-1.4\,GHz color) represents the
set of mock galaxies at all redshifts (drawn at random from the sample
of 20 templates) whose colors match that of the observed source.  The
galaxies used as templates are primarily in the local Universe, drawn
from a mix of normal star-forming galaxies to ultraluminous infrared
galaxies, plus a few $z\sim2$ quasars with cold-dust emission.  The
breadth of SEDs in the \citeauthor{aretxaga03a} work leads to broader
probability density distributions in redshift for any given source
because of the degeneracy between dust temperature and redshift.  This
implies that the uncertainty on the FIR-$z$ fitting technique is
naturally more realistic and representative of the true uncertainty of
the redshift constraint in comparison to a single template fit.

In the analysis of bright {\it Herschel} sources, \citet{ivison16a}
tests the applicability of seven different template SEDs against those
sources with confirmed CO-measured spectroscopic redshifts across the
range $1.5<z<6$.  These include four single-galaxy templates ---
SMM\,J2135-0102 at $z=2.3$ also known as the `Cosmic Eyelash'
\citep{swinbank10a}, Arp\,220 \citep{klaas97a,rigopoulou96a}, HFLS3 at
$z=6.34$ \citep{riechers13a}, and HATLAS\,J142413.9$+$022304, a lensed
system at $z=4.2$ also known as G15.141 \citep{cox11a} --- as well as
three composite SEDs of different DSFG samples from \citet{pope08a},
\citet{pearson13a} and \citet{swinbank14a}.  \citet{ivison16a} found
that only three of these templates were good estimators of redshift:
the Cosmic Eyelash, the \citet{pope08a} composite SED and the
\citet{swinbank14a} composite SED.  Of those three, the FIR-$z$
photometric redshift and its uncertainty is determined 
  using a maximum likelihood estimator ($e^{-\chi^2}$) with the
best-fit template.

\citet{brisbin17a} present another FIR-$z$ method that does not rely
on identification at 1.4\,GHz explicitly, but is based solely on the
FIR emission.  They use a sample of 16 DSFGs with known spectroscopic
redshifts, spanning $0.1<z<4.7$ with median $\langle z\rangle=2.2$.
They us a simple inverted parabolic fit to the photometric data to
infer the observed peak wavelength, $\lambda_{\rm observed\ peak}$.
They then fit a linear relationship between redshift and $\lambda_{\rm
  observed\ peak}$ to anchor the FIR-$z$ model.  The uncertainty in
the FIR-$z$ fit is derived from the scatter in the linear relation
between $z$ and $\lambda_{\rm observed\ peak}$.

The template SEDs often used in the literature are shown in
Figure~\ref{fig:sampleseds} relative to a generic SED peaking at a
rest-frame wavelength of $\lambda_{\rm peak}=100$\,\um.  In
\S~\ref{sec:tests}, I draw comparisons between the literature
approaches that are not anchored to any radio flux density
measurements, primarily the single galaxy templates presented by
\citet{ivison16a} and the unique approach of \citet{brisbin17a}.

\section{The ``\mmpz'' Fitting Technique}\label{sec:mmpz}

\subsection{Algorithm Design}

The \mmpz\ FIR-$z$ fitting technique is designed to provide an
accurate estimate of a galaxy's redshift based on its absolute FIR/mm
photometry (rather than relative photometry alone) {\it and} an
accurate representation of the uncertainty of that redshift estimate.
A good representation of the uncertainty relies on an understanding of
the intrinsic underlying variation of galaxies' dust SEDs.  Based on
hundreds of {\it Herschel}-observed galaxies from $0<z<3$, and the
South Pole Telescope sample of DSFGs observed toward higher redshifts
(out to $z\sim7$), \citet{casey18a} showed that galaxies' dust SEDs
follow a general trend relating the rest-frame peak wavelength \lpeak\
to the integrated IR luminosity \lir\ via:
\begin{equation}
\langle\lambda_{\rm peak}(L_{\rm IR})\rangle = \lambda_0
\bigg(\frac{L_{\rm IR}}{L_t}\bigg)^\eta
\label{eq:lirlpeak}
\end{equation}
Where $\lambda_0$=102.8$\pm$0.4\,\um, $L_t\equiv10^{12}$\,\lsun, and
$\eta=-0.068\pm0.001$.  This relation has a typical scatter in
$\log($\lpeak$)$ of $\sigma_{\log(\lambda)}=0.045$. This is given as
Eq.~2 in \citet{casey18a}.  Physically, this relationship implies that
more luminous galaxies have warmer luminosity-weighted dust
temperatures, which may be driven by a combination of harder radiation
fields from higher star formation rate densities and more compact dust
geometries.

\begin{figure*}
\centering
\includegraphics[width=0.99\columnwidth]{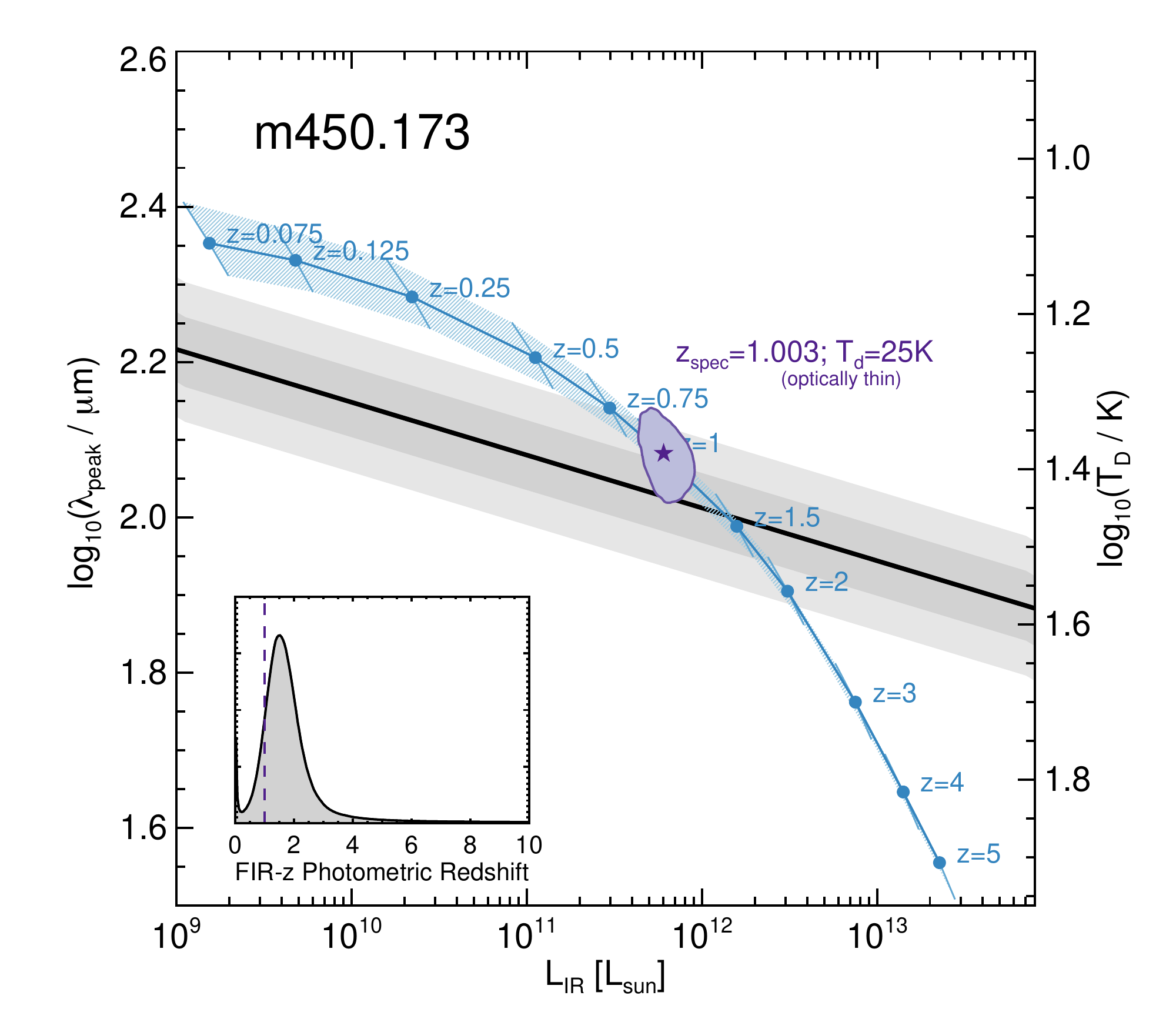}
\includegraphics[width=0.99\columnwidth]{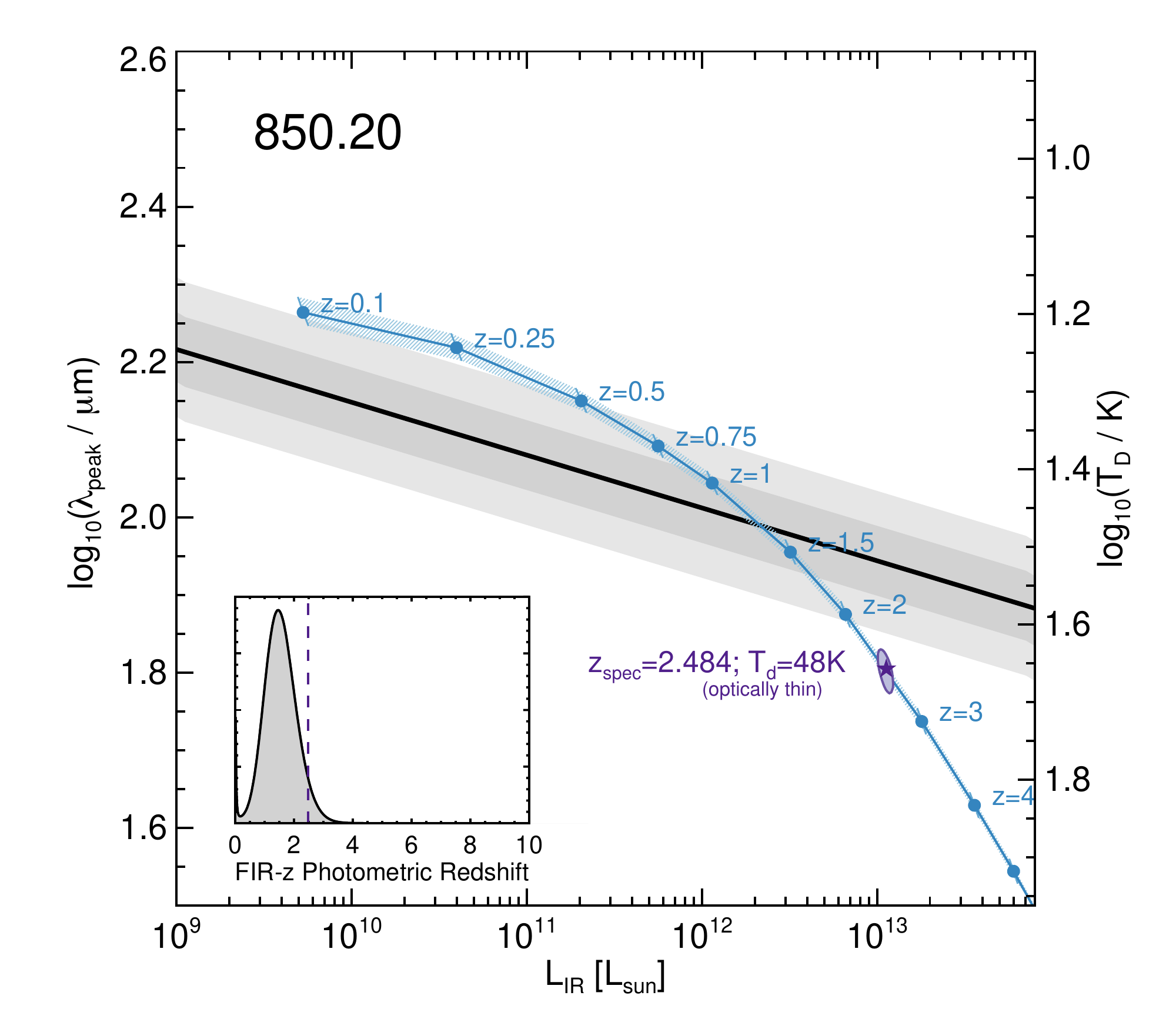}\\
\includegraphics[width=0.99\columnwidth]{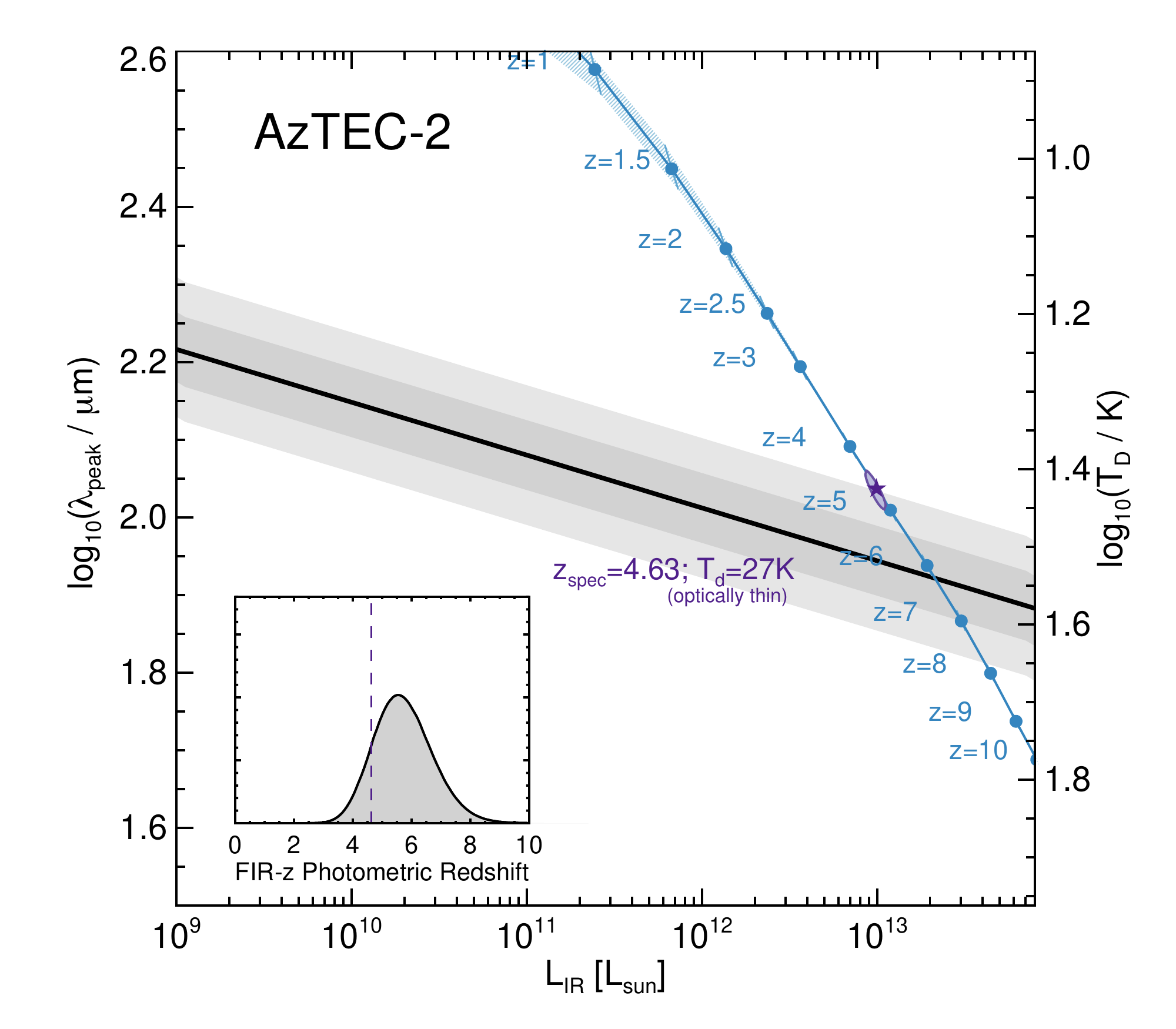}
\includegraphics[width=0.99\columnwidth]{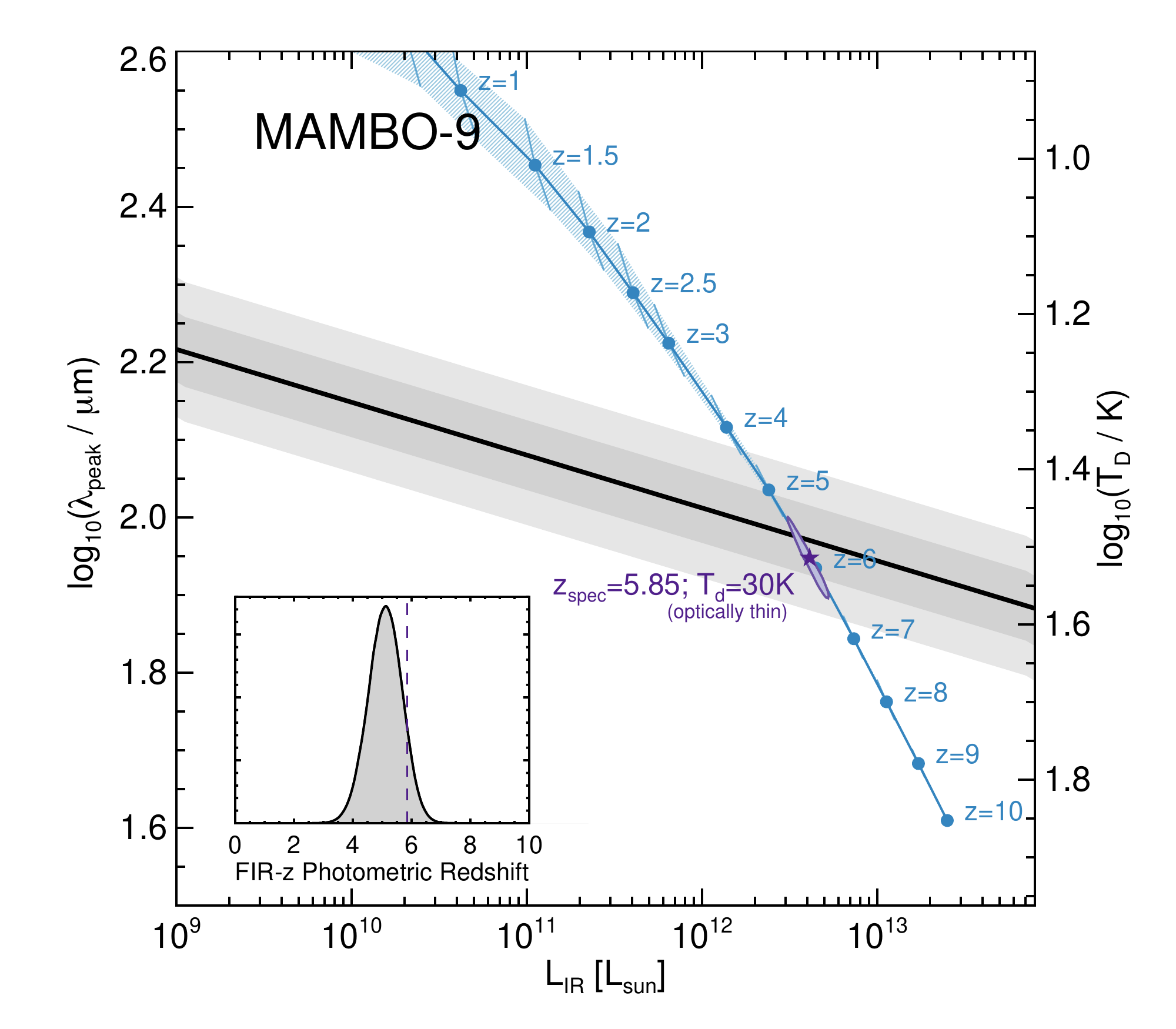}\\
\includegraphics[width=0.49\columnwidth]{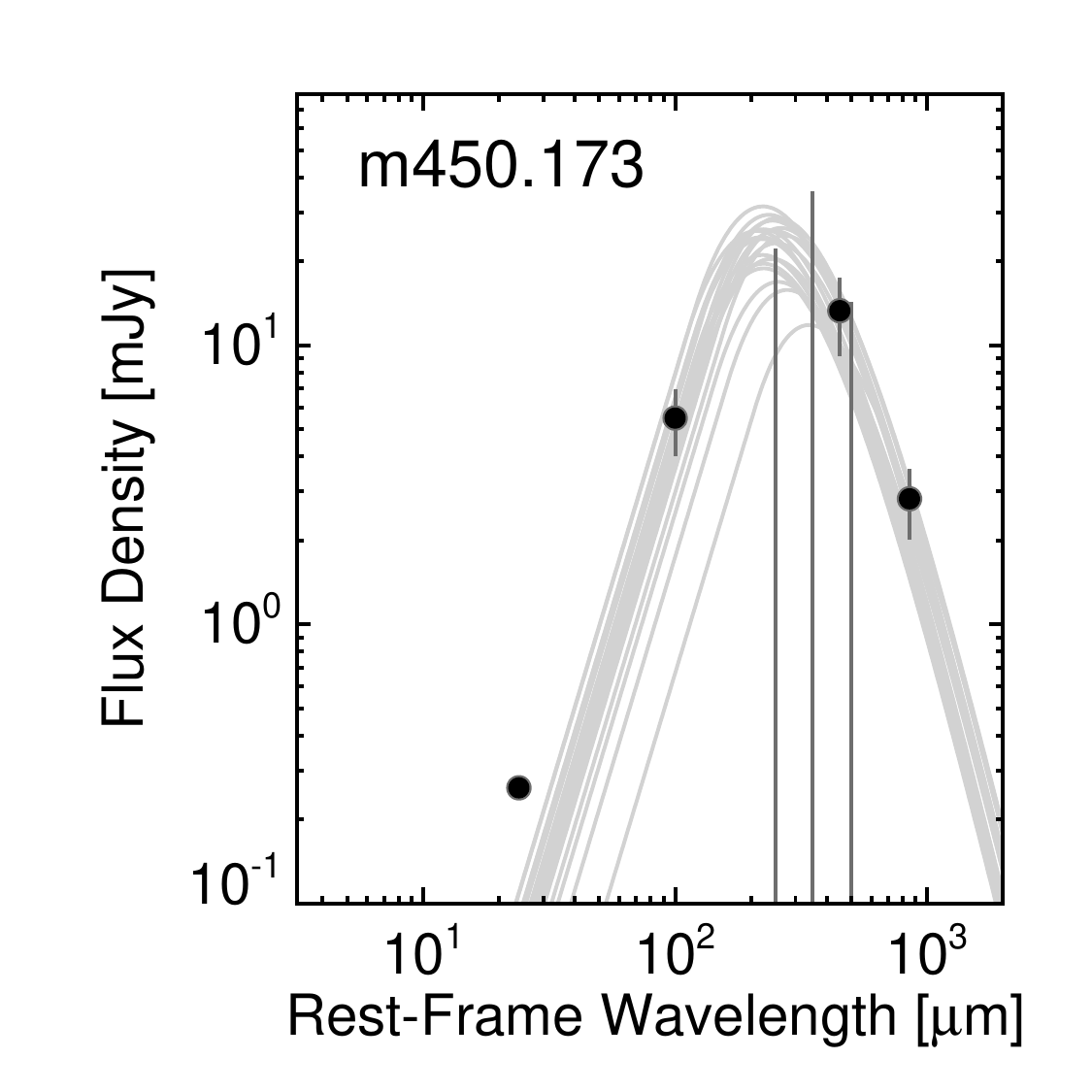}
\includegraphics[width=0.49\columnwidth]{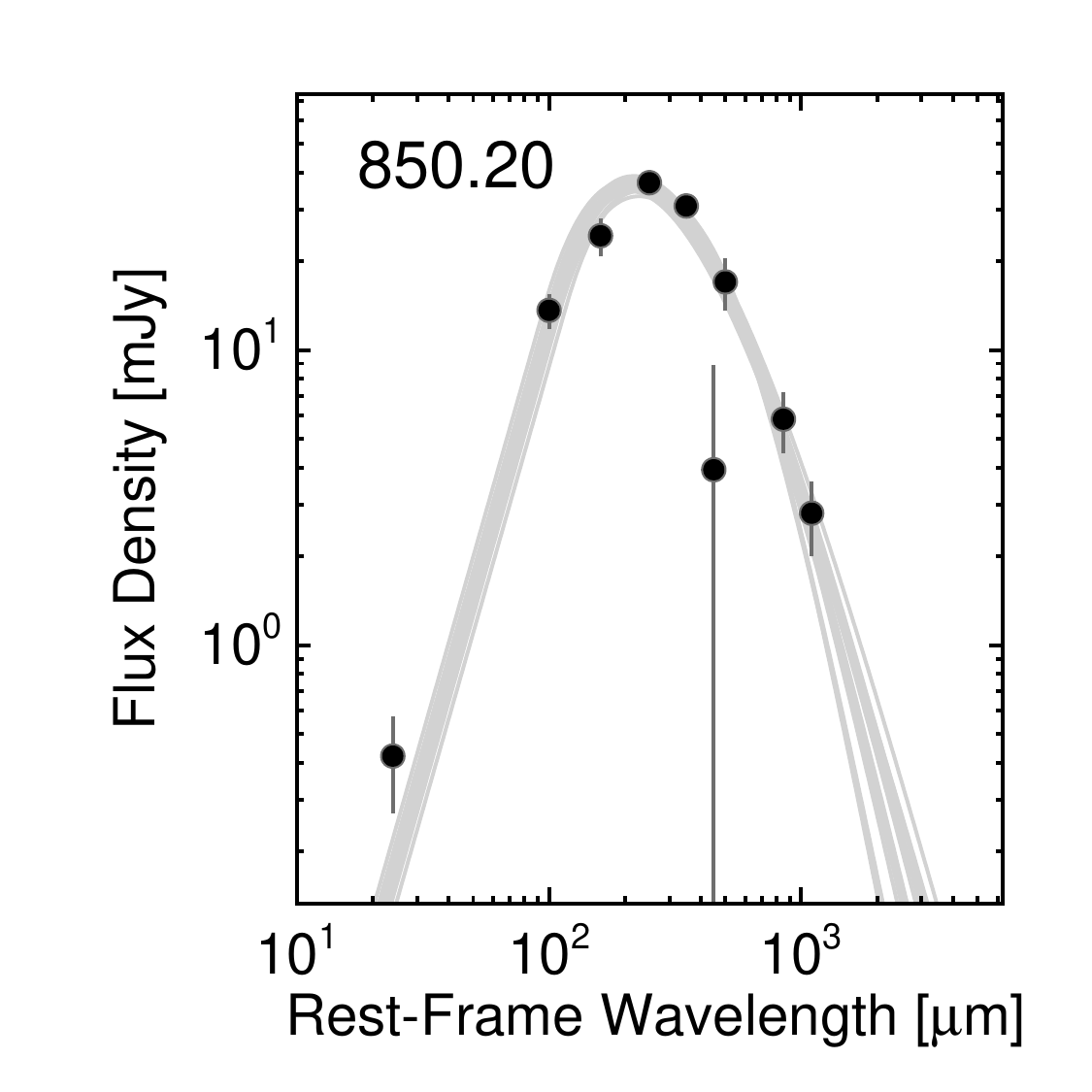}
\includegraphics[width=0.49\columnwidth]{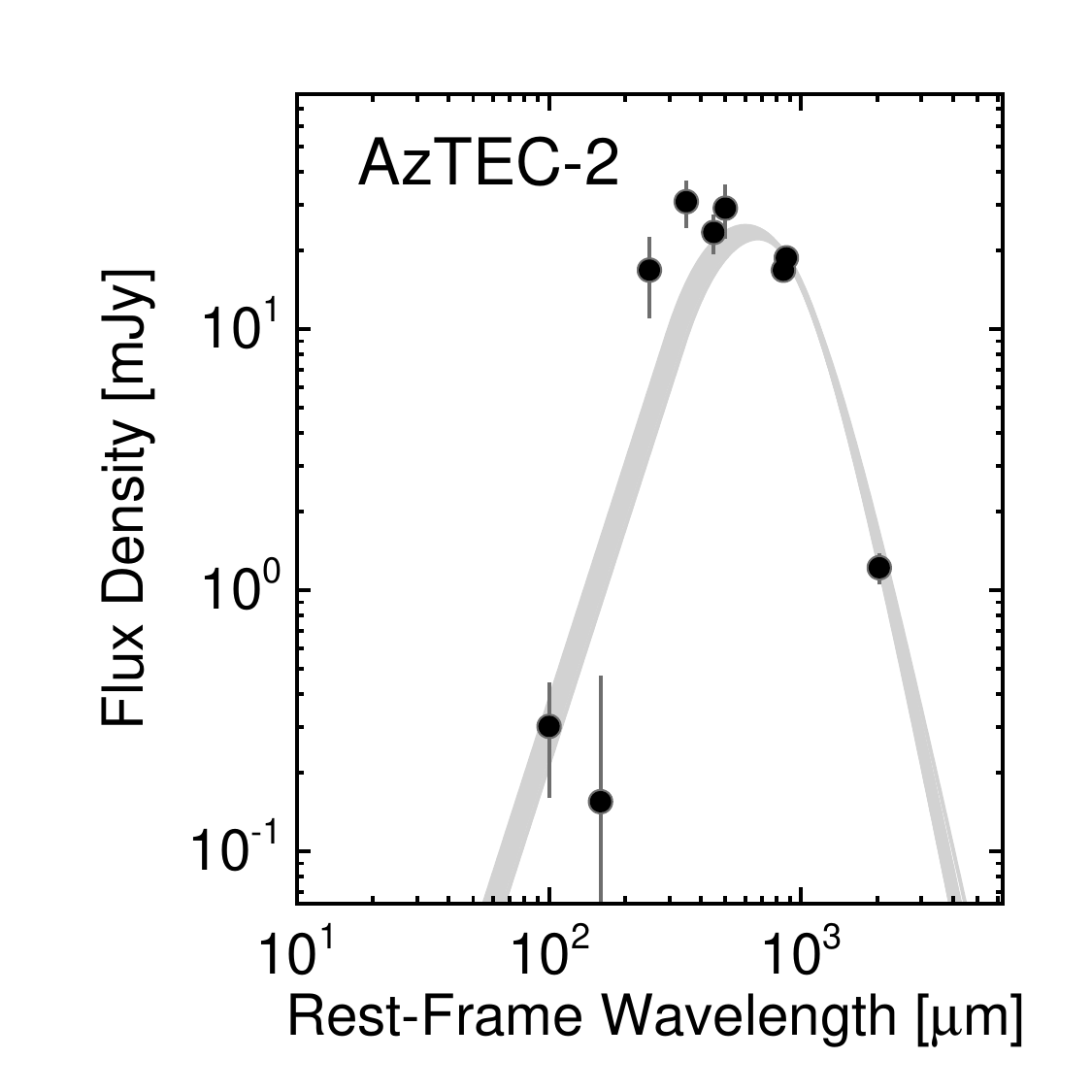}
\includegraphics[width=0.49\columnwidth]{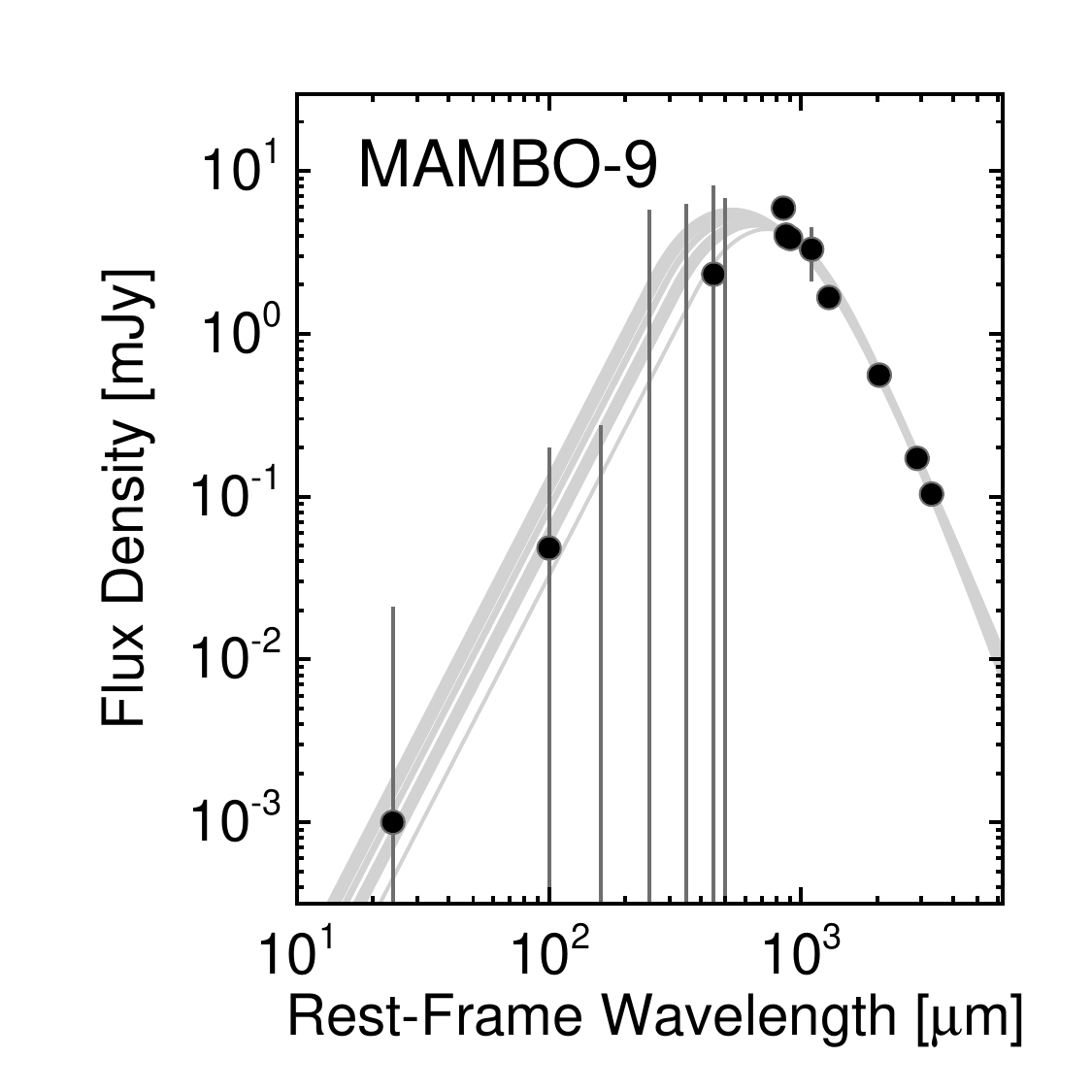}\\
\caption{A visual illustration of the \mmpz\ FIR-$z$ fitting technique
  using four spectroscopically-confirmed DSFGs at various redshifts:
  m450.173, a marginally-detected {\sc Scuba-2} 450\um\ and
  850\um\ source at $z=1.003$ \citep{casey17a}, 850.20, a
  850\um-detected {\it Herschel}-detected DSFG at $z=2.484$
  \citep{casey15a}, AzTEC-2, a 1.1\,mm-selected DSFG at $z=4.63$
  \citep[AzTEC2;][]{jimenez-andrade20a}, and MAMBO-9, a
  1-2\,mm-selected DSFG confirmed at $z=5.85$ \citep{casey19a,jin19a}.
  The set of photometric constraints for each galaxy (shown at bottom)
  traces out a limited track in the \lir-\lpeak\ plane, illustrated by
  the blue lines with associated uncertainty at each noted redshift
  represented by the thickness of the blue band.  The uncertainty from
  the photometry in \lir-\lpeak\ given the source's photometry {\it
    and} spectroscopic redshift is shown in the purple 1$\sigma$
  enclosed contour.  The empirical relationship between \lir\ and
  \lpeak\ is shown in the black line and associated 1$\sigma$ and
  2$\sigma$ scatter, shown in gray and light gray.  The probability
  density distribution in redshift is then constructed by sampling the
  gray shaded region (i.e. two dimensional probability distribution)
  along the blue redshift track (inset plots).}
  \label{fig:mambo9}
\end{figure*}

The luminosity dependence of the average peak wavelength is crucial
information that \mmpz\ folds into a galaxy's FIR-$z$ fit.  For
example, a given set of FIR/mm photometry suggests a limited intrinsic
range of IR luminosities, especially if that photometry is primarily
on the Rayleigh-Jeans tail of the peak, where flux density (and thus
\lir) is roughly constant with redshift due to the very negative
$K$-correction.  Thus, FIR-$z$ fitting should make direct use of flux
density measurements themselves, not just the contrast that is drawn
between bands through FIR/mm colors.  This is because sources with
intrinsically brighter flux densities are more likely to have higher
\lir, and thus more likely to have intrinsically warmer dust SEDs.
This strategy forms the backbone of the \mmpz\ technique.

The algorithm is illustrated using a few examples of DSFGs with known
redshifts in Figure~\ref{fig:mambo9}: MAMBO-9 at $z=5.85$
\citep{casey19a}, AzTEC-2 at $z=4.63$
\citep{jimenez-andrade20a}, 850.20 at $z=2.48$ \citep{casey15a}, and
m450.173 at $z=1.00$ \citep{casey17a}.  These systems are chosen
because they span a wide redshift range and wide range of intrinsic
SEDs. For all possible redshifts, the measured photometry constrain
the range of possible SEDs in the \lir\ and rest-frame \lpeak\ plane.
Each galaxy traces out a track in the \lir-\lpeak\ plane as a function
of redshift.  This track is shown in Figure~\ref{fig:mambo9} by the
blue lines and shaded region of uncertainty reflective of photometric
variance.  The shape and direction of the blue curve traces the nature
of the photometric constraints: an SED with measurements predominantly
on the Rayleigh-Jeans tail translates to vertical tracks in this
\lir-\lpeak\ plane while photometry constraining the peak is more
likely to result in uncertain \lir\ and better constrained \lpeak.
This is perhaps counterintuative, because galaxies that are well
sampled near their peaks should be relatively well constrained in both
\lir\ and \lpeak; however, if the redshift is unknown, this introduces
large uncertainty in \lir\ because peak constraints do not benefit
from the negative $K$-correction in the same fashion as do
Rayleigh-Jeans constraints.

The thickness of this blue shaded region at each marked redshift
traces out the $\pm$1$\sigma$ range of plausible SED solutions as
measured using Markov Chain Monte Carlo (MCMC)
modeling. We implement the MCMC fitting of the SED
  similarly to \citet{casey19a}, whereby the aggregate photometry is
  measured against SEDs with variable \lir\ and \lpeak\ at a fixed
  redshift, and the likelihood of accepting a given SED in an MCMC
  chain is proportional to $e^{-\chi^2}$.  Upper limits on photometric
  points are handled directly, whereby the algorithm takes as input
  both a flux density and uncertainty measurement, no matter the
  significance of the measurement; negative flux densities are also
  accepted as input (this can occur due to Gaussian fluctuations in a
  band with a non-detection).

Overall, the blue track in Figure~\ref{fig:mambo9}
  traces out the range of plausible \lir\ and \lpeak\ values
  constrained by photometry across a range of redshifts. The gray
region in Figure~\ref{fig:mambo9} centered on the black line traces
out the empirical relationship and scatter (1$\sigma$ and 2$\sigma$)
between \lpeak\ and \lir\ from measured data as stated above in
Equation~\ref{eq:lirlpeak}; in other words, 95\%\ of galaxies sit
within the outer bounded light gray region, regardless of redshift.

 The resulting \mmpz\ probability density distribution in redshift
 (i.e. the FIR-$z$ fit) is then generated by sampling the empirical
 \lir-\lpeak\ distribution (i.e. the two-dimensional gray distribution
 in Figure~\ref{fig:mambo9}) along the redshift track traced out by
 the source's photometry in \lir-\lpeak\ (i.e. the blue tracks in
 Figure~\ref{fig:mambo9}).  In the absence of a direct crossing of a
 source's redshift track and the mean \lir-\lpeak\ relationship
 (i.e. the black line in Figure~\ref{fig:mambo9}), the redshifts at
 which the two-dimensional probability distribution in \lir-\lpeak\ is
 maximized along the redshit track are designated the most likely.
 The further the redshift track is from the peak of the distribution
 in \lir-\lpeak, the less well-constrained the FIR-$z$ fit will be.

In practice, the probability density
  distribution in redshift is constructed by (1) using
a reference grid of SEDs populating the entire \lir-\lpeak\ plane
(described in the next subsection), (2) collapsing this
grid of SEDs to the wavebands where observations have been collected
for a given source; and then (3) comparing the grid
model photometry to the source's photometry using a
  $\chi^2$ maximum likelihood technique assuming Gaussian
uncertainties in the data; in other words, each grid
  point in the \lir-\lpeak\ plane has an associated $\chi^2$ with
  respect to the set of photometric constraints, and the likelihood is
  taken as $e^{-\chi^2}$ convolved with the probability density
  distribution of the given \lpeak\ as a function of \lir.  If a
source has fewer photometric data points available, then the range of
plausible SEDs in \lir-\lpeak\ space will be broader than if there are
many high quality photometric constraints.  However, a
well-constrained SED does not mean a well-constrained FIR-$z$ fit, due
to the degeneracy of dust temperature and redshift with the observed
peak wavelength, and relatively broad intrinsic range of SEDs even at
a fixed \lir.

\subsection{Choice of SED shape and Input Information}

The reference grid of SEDs used in \mmpz\ is generated using a simple
modified blackbody and mid-infrared powerlaw model as in
\citet{casey12a}, where $\alpha_{\rm MIR}=3$ represents the
mid-infrared powerlaw slope and $\beta=1.8$ is the emissivity spectral
index.  Instead of using the analytic approximation as in
\citet{casey12a}, we use a piece-wise function
\footnote{ The reason for the shift to a piece-wise
    function is only aesthetic; the analytical form has a small `bump'
    visible in the SED when plotted in log-wavelength vs log-flux
    density.  This bump has no impact on the measured \lir\ or
    \lpeak\ of a derived fit, but is not well motivated physically.
    The piece-wise function more accurately reflects the underlying
    physical model of a smooth powerlaw distribution in dust
    temperatures.}
 for the modified
blackbody plus mid-infrared powerlaw as in
\citet{casey19a}.

What is the impact of these fixed values for $\alpha_{\rm MIR}$ and
$\beta$ on the FIR-$z$ fit? A range of physical values for both
parameters were tested for their impact on the output FIR-$z$ results.
The probability density distribution in $z$ is completely insensitive
to $\beta$ (given the negligible contribution of light on the
Rayleigh-Jeans tail of the blackbody to \lir).  The mid-infrared
$\alpha_{\rm MIR}$ slope has more impact.  While real galaxies have a
variety of mid-infrared powerlaw slopes, with lower values of
$\alpha_{\rm MIR}\sim1$ corresponding to more mid-infrared emission
(perhaps hinting at hot dust surrounding an active galactic nucleus,
AGN), the vast majority of DSFGs have moderately steep mid-infrared
slopes, $2\simgt\alpha_{\rm MIR}\simgt5$.  As $\alpha_{\rm MIR}$
increases, the mid-infrared component of the SED becomes negligible,
contributing $\simlt$5\%\ to the total \lir.  Note that the DSFGs
requiring a FIR-$z$ fit tend to have fewer rest-frame mid-infrared
constraints than DSFGs that have alternate redshift estimators; the
presence of mid-infrared counterparts implies that a
optical/near-infrared counterpart is more likely to exist
\citep{magdis12a}, leading to an OIR constraint on the redshift.
FIR-$z$ fits generated from data primarily on the Rayleigh-Jeans tail
of blackbody emission are thus not impacted by the choice of
$\alpha_{\rm MIR}$.  In summary, we determine that our choice of
$\alpha_{\rm MIR}=3$ and $\beta=1.8$ have minimal impact on the
resulting FIR-$z$ fits from \mmpz.

One important effect that the reference grid of SEDs takes into
consideration is heating from the Cosmic Microwave Background at
high-redshifts \citep[at $z\simgt5$;][]{da-cunha13a}.  The CMB
heats the ISM of high-$z$ galaxies, which in turn diminishes the
contrast between the galaxy and its background, thus directly
impacting the measured flux densities of dusty galaxies at
sufficiently high redshifts (where the CMB temperature was much higher
than it is today).  The modeling of the impact of the CMB in the
\mmpz\ grid construction follows the prescription in
\citet{da-cunha13a}.

As input, the \mmpz\ technique requires at least two
  flux density constraints in the FIR/mm.  Non-detections can (and
  should) be included by reporting a measured flux density at a
  corresponding wavelength with associated flux density uncertainty.
  In practice, the list of photometric input datapoints can be an
  amalgamation of data from both single-dish telescopes and
  interferometers, including data with dramatically different
  beamsizes, sensitivities, etc.  The user should make every attempt
  to reconstruct the intrinsic flux density of the given source from
  the given measured photometry, whether or not that includes
  accounting for a deboosting factor.  Section~\ref{sec:spire} issues
  a cautionary note with respect to {\it Herschel}-SPIRE flux
  densities in particular, but the user should be aware that the
  \mmpz\ algorithm itself will not directly account for any effects of
  confusion or Eddington boosting based on the instrument from which
  the data originates.

One subtle point to note is that the \mmpz\ technique
  does not intrinsically account for the bandpass sensitivity curves
  of various FIR/mm instruments.  Instead, it assumes reported flux
  densities are equal to the flux density intrinsic to the underlying
  SED.  This is a common practice for fitting relatively simple SEDs
  to relatively sparse data at long wavelengths.  This approximation
  greatly simplifies the computational burden of the algorithm, and is
  appropriate when the stated wavelength of observations is equal to
  the isophotal wavelength corresponding to the set of observations
  \citep[e.g. as explicitly defined in][]{tokunaga05a}. In the FIR/mm,
  the differences between measuring flux densities as a convolution of
  a filter sensitivity curve and an underlying SED versus
  approximating flux density from the SED at a fixed wavelength
  amounts to a 1-3\%\ effect, which is nearly always negligible
  relative to the total uncertainty in the flux density measurements.

Once a probability density distribution in redshift for a given source
is in-hand, the optimum FIR photometric redshift FIR-$z$ is determined
by taking the mode of the distribution and a 68\%\ minimum credibility
interval, or the minimum interval in redshift over which 68\%\ of the
distribution is contained.  Adopting the mode of the distribution as
the optimum redshift rather than the median redshift is appropriate in
particular for bimodal or asymmetric probability distributions, which
does occur fairly frequently using the \mmpz\ technique.

\subsection{A Cautionary note about {\it Herschel}-SPIRE}\label{sec:spire}

The {\it Herschel Space Observatory} SPIRE instrument
  has provided a wealth of data from 250--500\um\ across several
  extragalactic survey fields and over large swaths of the
  extragalactic sky \citep[e.g.][]{Oliver12a}.  This wavelength range
  can be crucial to constraining the peak of the SED for $z>1$ DSFGs.
  However, given the large beamsize of observations (18$''$--36$''$)
  and high on-sky source density, SPIRE maps are highly confused.  The
  measured mean point-source confusion noise for SPIRE is 5.8, 6.3,
  and 6.8\,mJy/beam at 250, 350, and 500\um, respectively
  \citep{nguyen10a}; the confusion noise dominates over the
  instrumental noise in most SPIRE maps (with the exception being the
  shallowest SPIRE maps).  Sophisticated deblending algorithms have
  attempted to use positional priors (e.g. from radio continuum,
  24\um\ or other wavelengths where positions of sources are
  well-constrained) to extract and accurately measure flux densities
  below this confusion limit \citep{roseboom10a,liu18a,jin18a}.  While
  the technique of using positional priors is innovative and can be
  useful for characterizing populations of galaxies in aggregate, it
  is problematic when used to constrain the SED of an individual
  source that isn't detected at high significance.  In particular,
  these deblending algorithms are prone to under-estimating the
  uncertainty on a SPIRE flux density measurement dramatically, as
  they cannot account for uncharacterized uncertainty in the
  underlying model.  Because the SPIRE data can have such a
  substantial impact on the shape of an SED fit despite this
  uncertainty, I caution users of the \mmpz\ algorithm (and more
  broadly) to use the measured point-source confusion noise estimates
  as a lower limit on the flux density uncertainties for SPIRE.

\section{Comparison between {\sc mmpz} and other techniques}\label{sec:tests}

In this section I compare the \mmpz\ FIR-$z$ technique to other
radio-independent FIR-$z$ fitting techniques by generating SEDs of
mock galaxies of known redshifts and use them to assess the accuracy
and precision of each FIR-$z$ technique.  In addition, a smaller
sample of spectroscopically-confirmed real DSFGs is also used to
compare methods as a check on the results generated from much larger
samples of mock galaxies.

\begin{figure*}
\centering
\includegraphics[width=0.99\columnwidth]{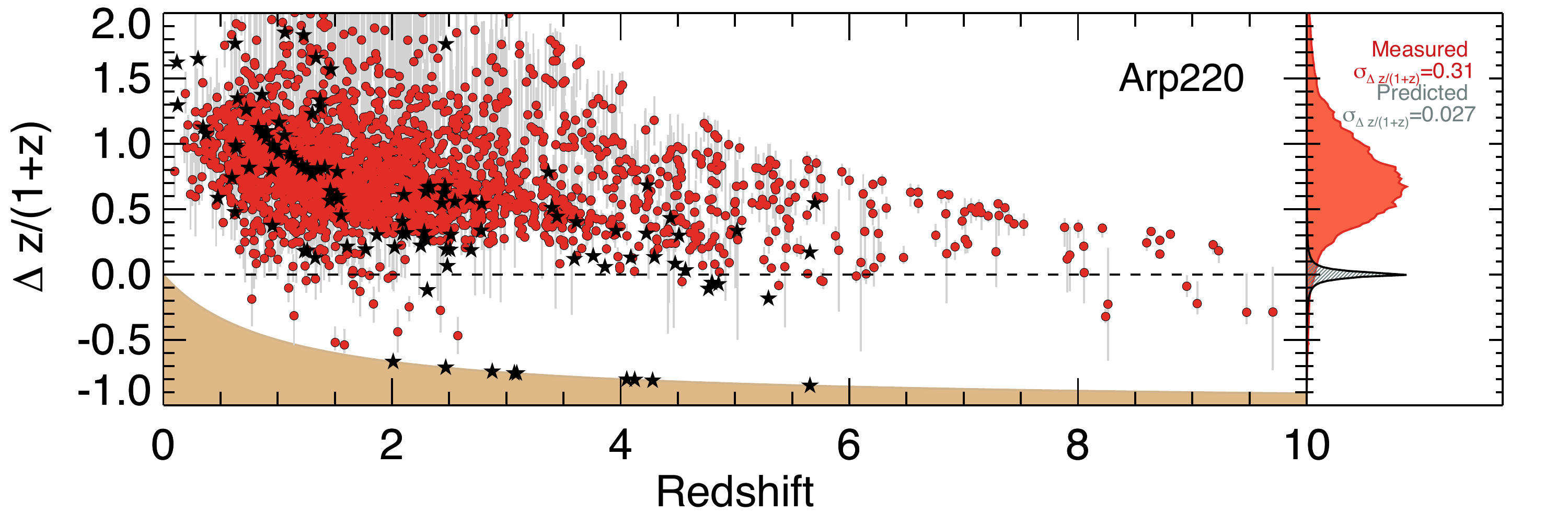}\includegraphics[width=0.99\columnwidth]{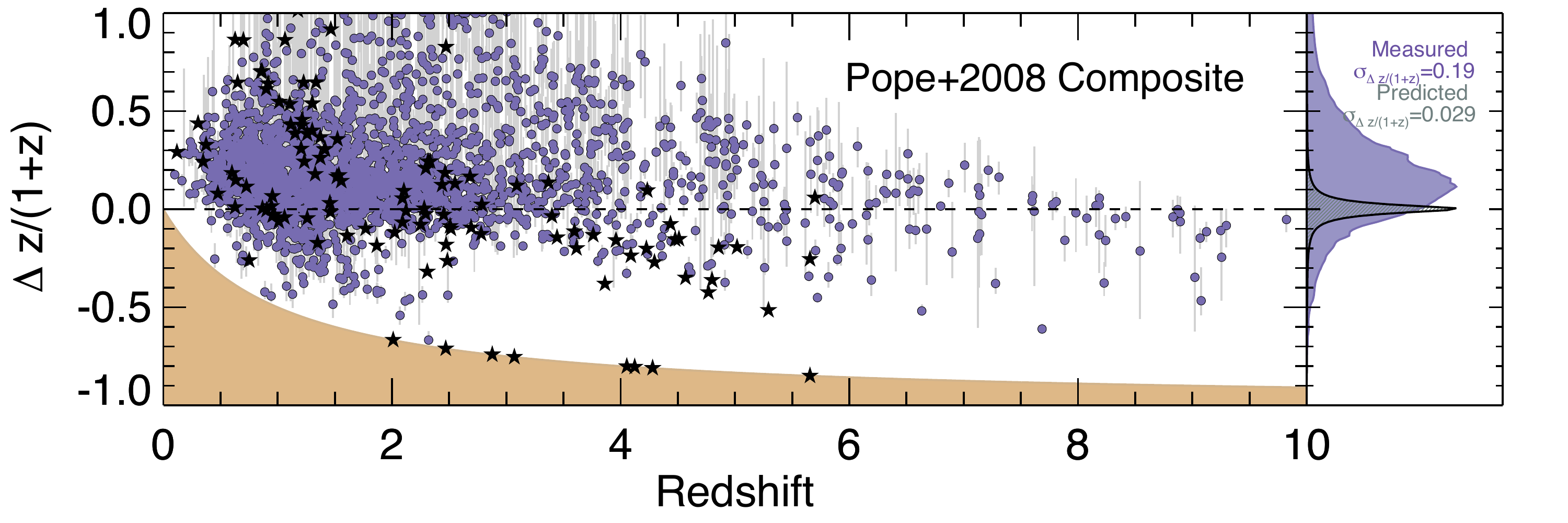}\\
\includegraphics[width=0.99\columnwidth]{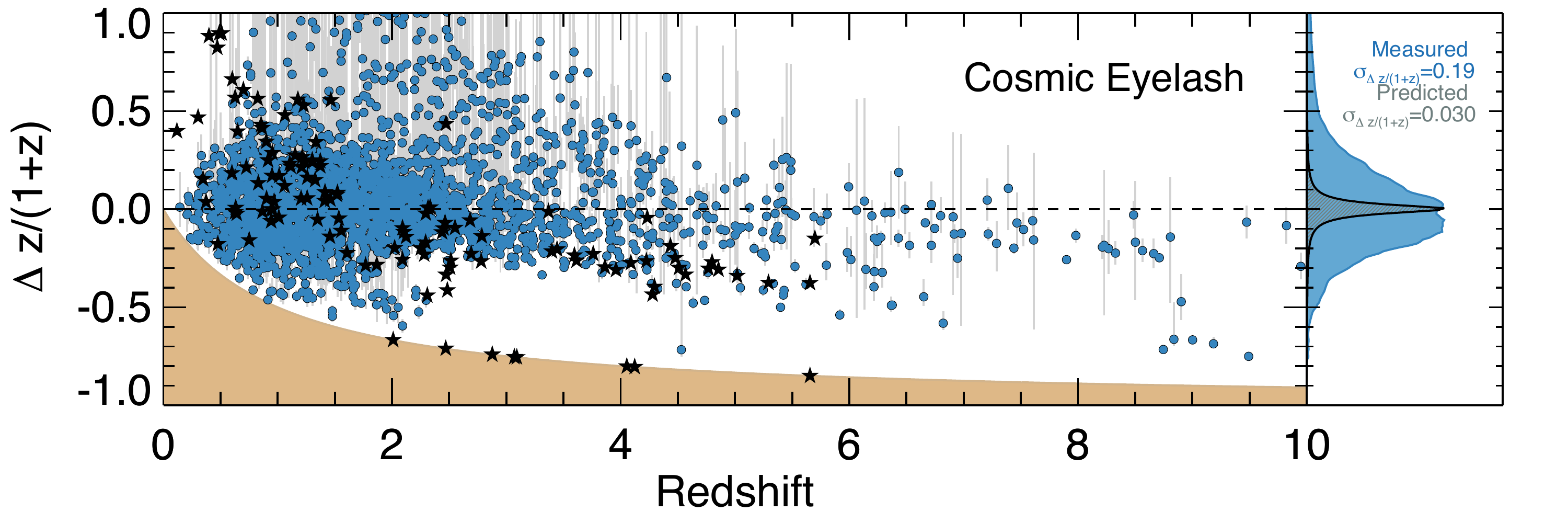}\includegraphics[width=0.99\columnwidth]{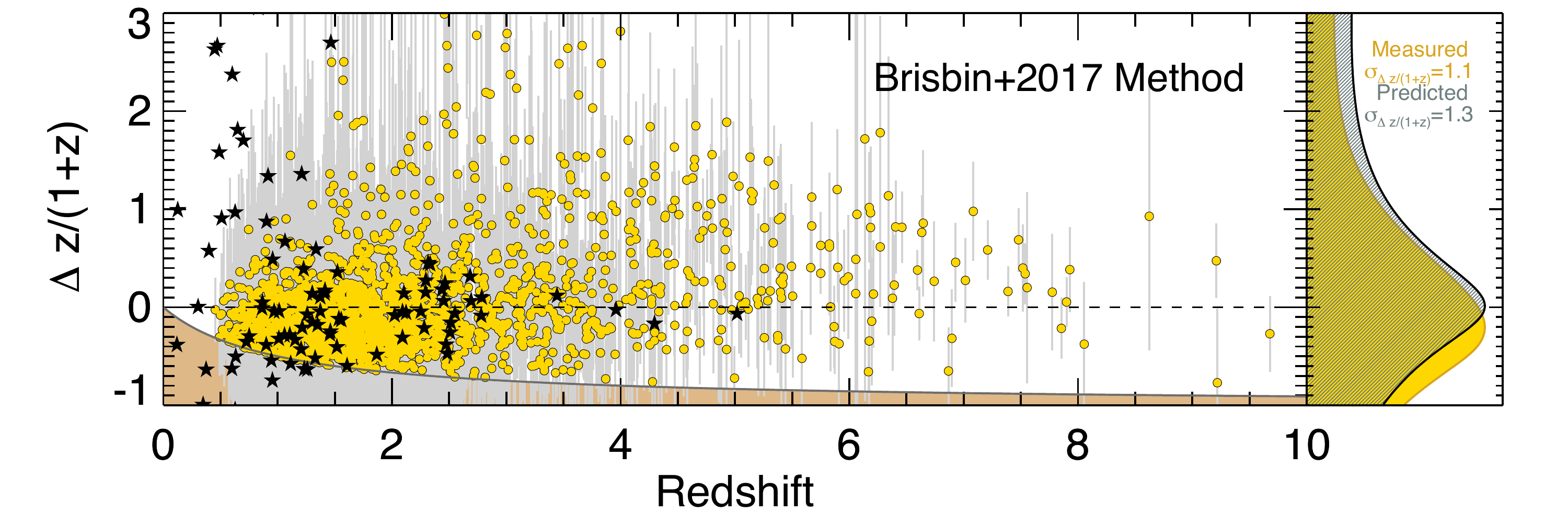}\\
\includegraphics[width=0.99\columnwidth]{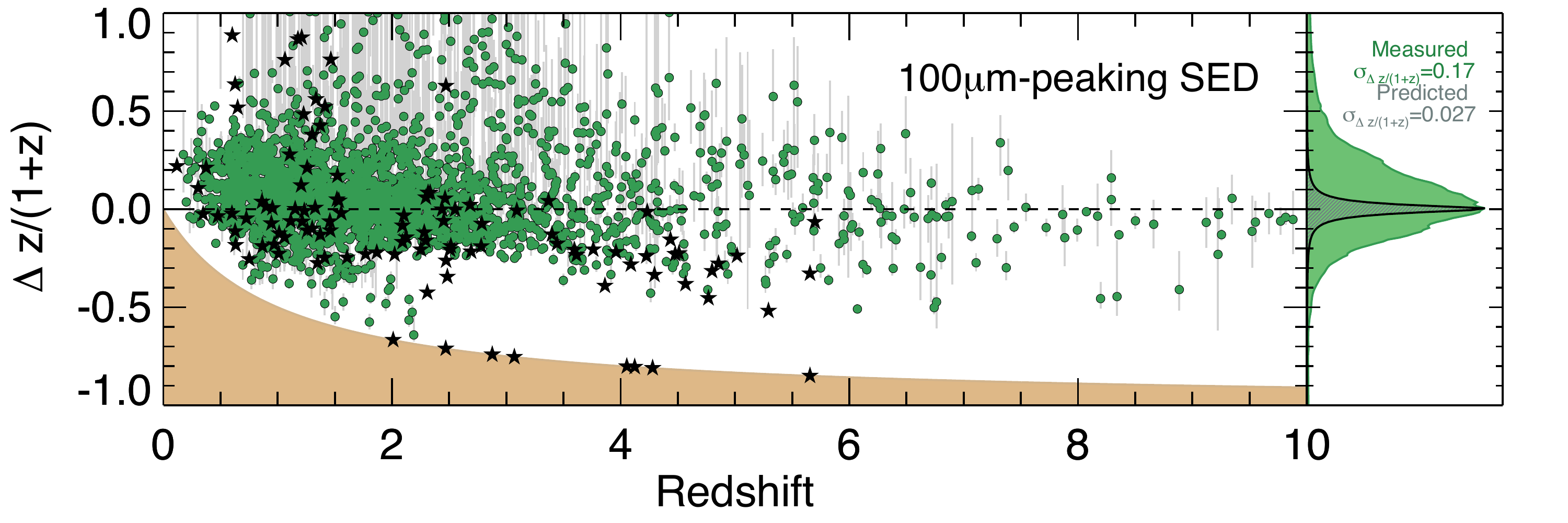}
\includegraphics[width=0.99\columnwidth]{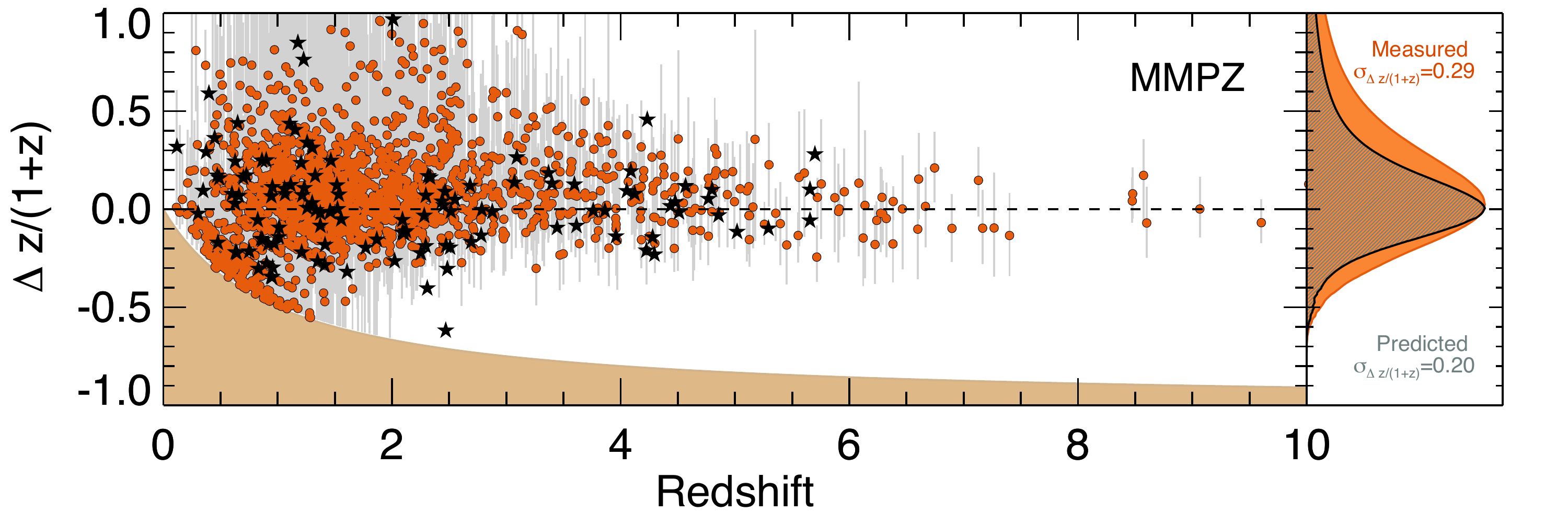}\\
\caption{Redshift against $\Delta z/(1+z)$ for mock galaxies described
  in \S~\ref{sec:tests}. Values of $\Delta z/(1+z)>0$ indicate
  photometric redshifts higher than the mock galaxy's true redshift,
  while values less than 0 are fit at lower redshifts.  The solid tan
  region in each main panel shows excluded parameter space
  (i.e. corresponding to $z<0$, or blueshifted galaxies).  Four panels
  represent single-template FIR-$z$ estimates: Arp220 (red, top left),
  the Cosmic Eyelash (blue, middle left), a 100\um-peaking SED (green,
  lower left), and the \citet{pope08a} composite (purple, top right).
  The last two panels represent FIR-$z$ fits from the method outlined
  in \citet[][yellow, middle right]{brisbin17a} and the
  \mmpz\ technique (orange, bottom right).  The right-most panel of
  each plot shows the collapsed, coadded probability density
  distributions in $\Delta z/(1+z)$ for the aggregate measurements of
  the mock galaxy population (solid colored histograms).  The black
  hashed histogram denotes the {\it predicted} collapsed probability
  density distributions in $\Delta z/(1+z)$, if a mock galaxy's
  estimated redshift is adopted as its true redshift.  The
  single-template SED fits have much broader distributions in $\Delta
  z/(1+z)$ than would be predicted from the method's reported
  uncertainties.  The uncertainties in the \citet{brisbin17a} method
  and the \mmpz\ technique are well matched to the overall
  distribution in $\Delta z/(1+z)$, indicating both provide a better
  estimate of precision than single-template fits.  Of these, the
  \mmpz\ technique provides better overall accuracy.}
\label{fig:simple}
\end{figure*}

Mock galaxies are drawn from the observed distribution of galaxies in
the \lir-\lpeak\ plane following Equation~\ref{eq:lirlpeak} (with
associated scatter), and following the best current estimate of the
evolving IRLF as given in \citet{zavala18a}.  Thousands are simulated
over a very large volume so that the technique can be tested over a
wide range of redshifts and dynamic range in luminosity.  Each mock
galaxy is then downsampled in wavelength space to observing bands one
might have on-hand for DSFGs in the literature, and Gaussian noise is
added according to the typical RMS of observations in those bands with
those instruments.  These bands include {\it Spitzer} 24\um; {\it
  Herschel}-PACS 70\um, 100\um, and 160\um; {\it Herschel}-SPIRE
250\um, 350\um, and 500\um; {\sc Scuba-2} 450\um\ and 850\um; AzTEC
1.1\,mm; GISMO 2.0\,mm; and ALMA bands 4--7 (from 870\um--2\,mm).  The
typical uncertainties assumed for these bands are given in Table~1 of
\citet{casey18a} and Table~1 of \citet{casey18b}.
It should be noted that many real galaxies will not have such a
plethora of SED constraints as modeled here, particularly if they sit
in regions of the sky not sampled by {\it Herschel}, or if fewer
high-precision measurements exist from ALMA.  Those galaxies will have
more uncertain FIR-$z$ fits.  The SED constraints that are modeled
here are representative of constraints that are likely to exist for
high-$z$ galaxies in extragalactic legacy fields (like COSMOS,
UDS/CDF-S, SXDF, etc.).  While our mock sample constrains many
intrinsically faint systems, we require at least three photometric
constraints above $>$3$\sigma$ significance for inclusion in our
analysis, as galaxies falling below the detection thresholds will have
no viable redshift constraints.

The spectroscopically-confirmed sample of DSFGs used to compare these
FIR-$z$ fitting methods come from three parent samples: the {\it
  Herschel}-SPIRE selected sample of spectroscopically-confirmed DSFGs
at $z\simlt1.5$ from \citet{casey12b}, 13 ALESS 870\um-selected DSFGs
from $1.5<z<2.5$ from \citet{swinbank14a} and \citet{danielson17a}, 18
SCUBA-2 spectroscopically-confirmed DSFGs from $0.5\simlt z\simlt2.6$
from \citet{casey17a} and the SPT-selected sample of
lensed DSFGs from \citet{weis13a}, with subsequent characteristics
described in \citet{strandet16a}, \citet{spilker16a} and
  \citet{reuter20a}.  The {\it Herschel} sample is down-sampled to
only include 62 galaxies detected at $>$4$\sigma$ in all three SPIRE
bands, as galaxies with less than this have poorly constrained SEDs.
The SPT sample photometry is corrected for gravitational lensing, as
using uncorrected photometry would skew the results of the \mmpz\ fits
which depend on flux density as well as colors; 36 galaxies from the
SPT sample have both well-measured magnification factors, $\mu$, and
spectroscopic redshifts.  In total, there are 129 galaxies in the
spectroscopic sample spanning spectroscopic redshifts $0.1<z_{\rm
  spec}<5.7$.

\begin{figure*}
\includegraphics[width=0.99\columnwidth]{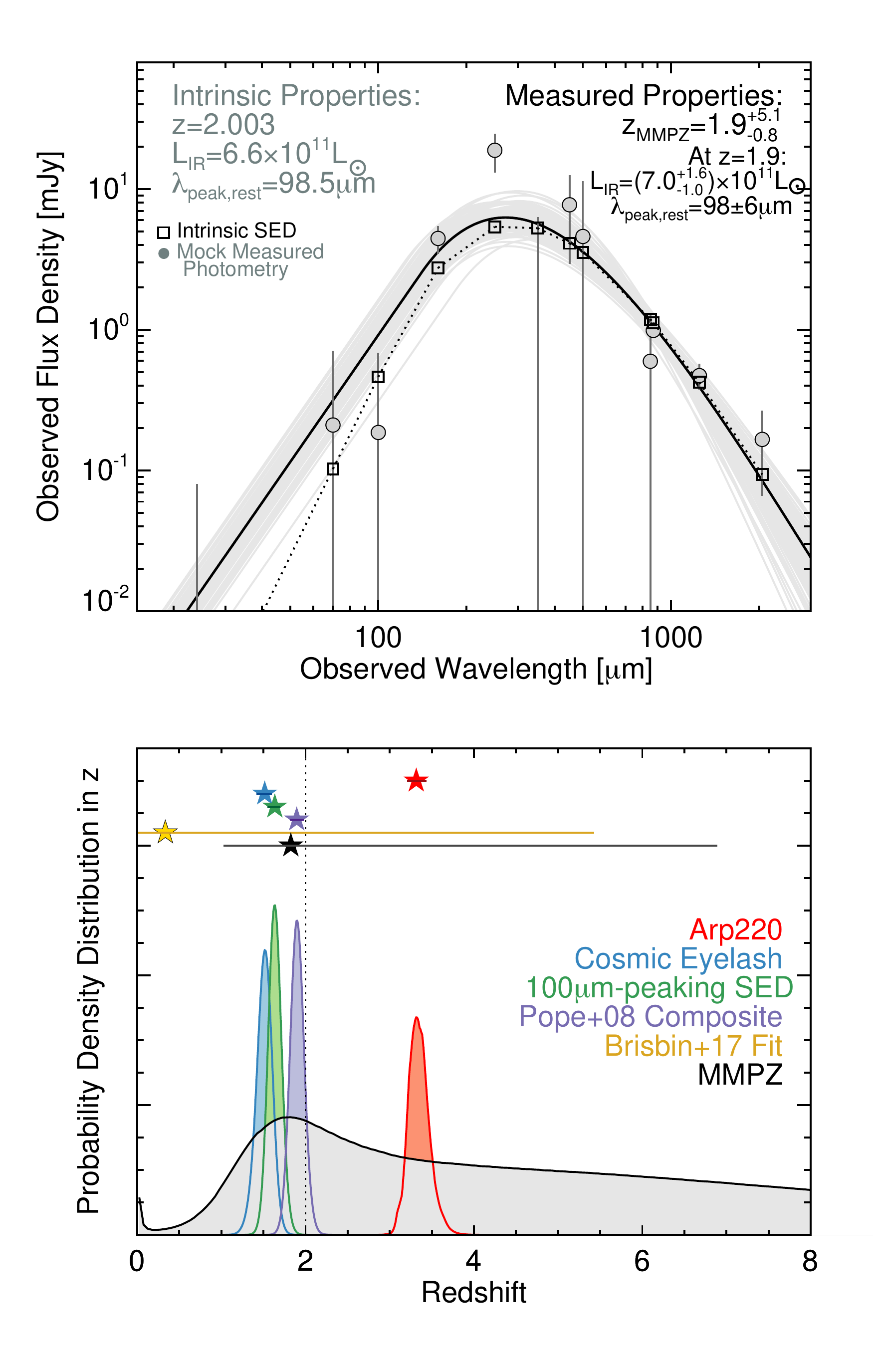}
\includegraphics[width=0.99\columnwidth]{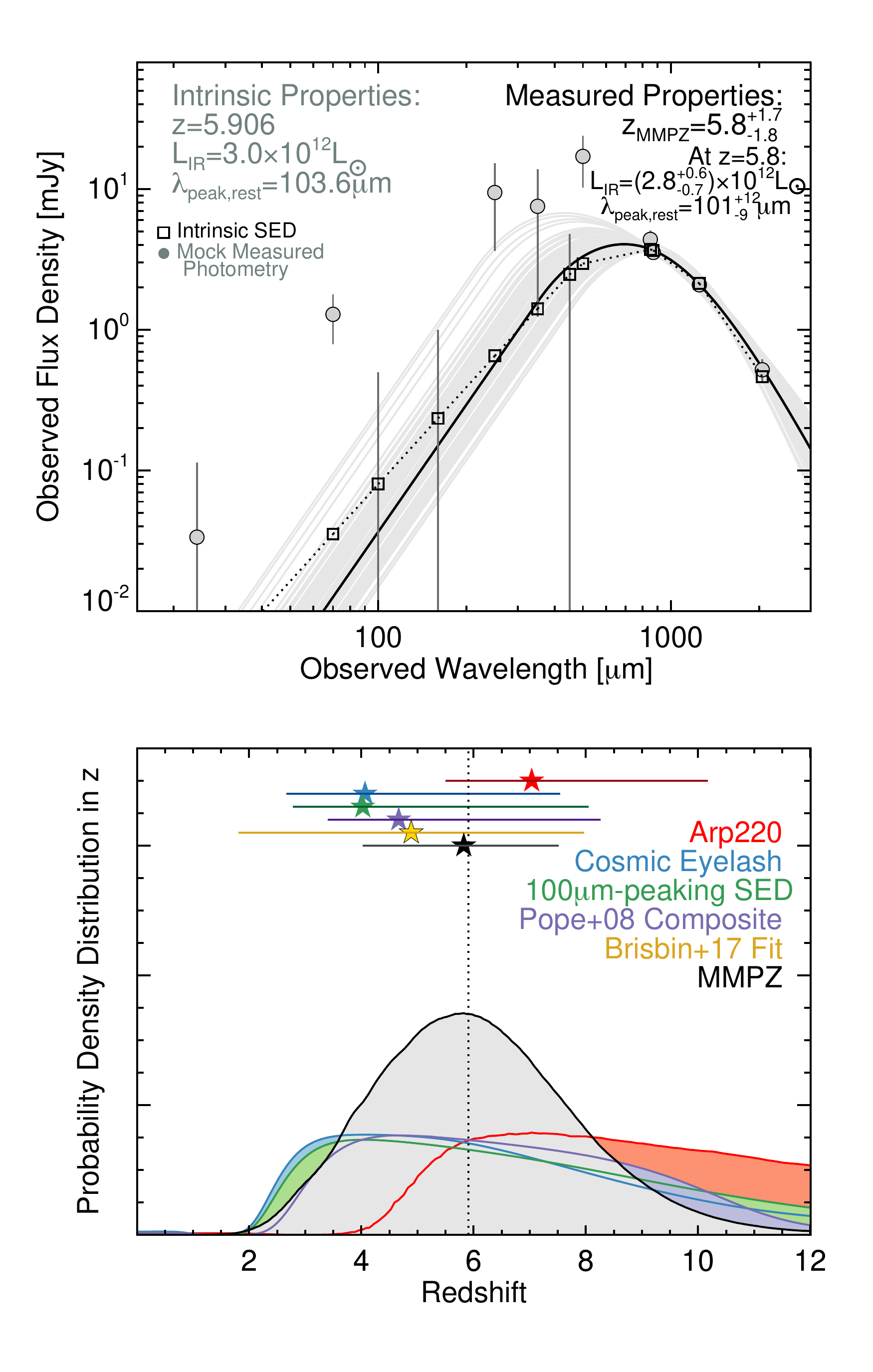}
\caption{Example mock galaxies' SEDs (top panels) and their resulting
  FIR photometric redshift probability density distributions (bottom
  panels).  The photometry from the intrinsic underlying SED is shown
  as black boxed points connected with a dotted line.  Gaussian noise
  is added to each point according to the typical noise of
  observations in each band, and the photometric redshift fits are run
  on the noise-added photometry (gray circles).  The overplotted gray
  SEDs represent draws from Markov Chain Monte Carlo trials fit to the
  mode of the \mmpz\ redshift distribution, and the black solid SED is
  the median SED of those trials.  The galaxies' intrinsic properties
  and measured properties are quoted in the inset; the measured
  \lir\ and \lpeak\ are quoted with fixed redshift.  The bottom panels
  show the probability density distributions of various fitting
  methods (following the same color scheme as in
  Figure~\ref{fig:simple}; no probability density function is given
  for the \citealt{brisbin17a} fits).  Note that the $z\sim2$ mock
  galaxy has a highly uncertain FIR-$z$ fit from \mmpz\ due to the low
  signal-to-noise of many of the flux density constraints (and low
  \lir).}
\label{fig:mock}
\end{figure*}

Figure~\ref{fig:simple} shows the accuracy and estimated precision of
six FIR-$z$ techniques: four single SED fits (using an Arp220 template
in red, the Cosmic Eyelash in blue, a 100\um-peaking SED in green, and
the \citealt{pope08a} composite SED in purple), the \citet{brisbin17a}
technique in yellow and the \mmpz\ technique in orange.  The resulting
FIR-$z$ fits for the spectroscopic sample are shown as black stars on
each panel.  Results from the 100\um-peaking SED are virtually
identical to results using the \citet{swinbank14a} composite SED so we
only show one panel.
The techniques that make use of a single-template SED vary in
accuracy.  The Arp220 SED, for example, seems to fit FIR-$z$ solutions
skewed toward higher redshifts than the intrinsic simulated mock
redshifts and spectroscopic redshifts.  The \citet{pope08a} composite
spectrum also skews toward higher redshifts, whereas the Cosmic
Eyelash and the 100\um-peaking SED fit fairly accurate SEDs for the
overall distribution of galaxies.  The primary shortcoming of these
single-template fits, however, is the estimated precision in the
FIR-$z$.  The hashed gray distribution in each right histogram panel
of Figure~\ref{fig:simple} highlights the distribution of
uncertainties estimated using the given technique.  A narrow
distribution compared to the overall distribution of fits in $\Delta
z/(1+z)$ (solid filled color histograms), represents a fit whose
uncertainty has been dramatically under-estimated.  This is the case
for all single-template SED fits because they cannot account for the
additional uncertainty brought on by a natural variance in galaxies'
SEDs.

In contrast to single-template fits, the \citet{brisbin17a} and
\mmpz\ techniques have broad distributions for the predicted
uncertainties.  Because the predicted uncertainties are similarly
broad to the overall distribution in $\Delta z/(1+z)$ for the entire
population (i.e. the hashed gray and solid histograms in
Figure~\ref{fig:simple} are similar), both the \citet{brisbin17a} fits
and the \mmpz\ fits are far more likely to capture the precision to
which constraints can be made from the FIR/mm photometry.  Between the
two fitting methods, the \mmpz\ provides a much tighter and more
accurate distribution in $\Delta z/(1+z)$.

\begin{table*}
\centering
\caption{Summary of Goodness-of-Fit to FIR-$z$ Methods.}
\begin{tabular}{c|ccccc|ccccc}
\hline\hline
{\sc Method} & \multicolumn{5}{c}{\underline{\sc Mock Galaxies}} & \multicolumn{5}{c}{\underline{\sc Real Spec-$z$ Confirmed Galaxies}}\\
  & $\langle {\Delta z/(1+z)}\rangle $ & $\sigma_{\Delta z/(1+z)}$ & $\chi^2$ & $\nu$ & $\chi^2_{\nu}$ & $\langle {\Delta z/(1+z)}\rangle $ & $\sigma_{\Delta z/(1+z)}$ & $\chi^2$ & $\nu$ & $\chi^2_{\nu}$ \\
\hline
Arp220             & 0.71    & 0.31 & $1.6\times10^{6}$ & 2498 & 634 & 0.64 & 1.07 & 2.4$\times10^{5}$ & 128 & 1860 \\
Cosmic Eyelash     & $-$0.02 & 0.19 & $1.1\times10^{5}$ & 2515 & 43.6 & $-$0.02 & 0.43 & 5.9$\times10^{4}$ & 128 & 459 \\
100\um-peaking SED & 0.04    & 0.17 & $1.1\times10^{5}$ & 2434 & 43.7 & $-$0.06 & 0.64 & 6.1$\times10^{4}$ & 128 & 475 \\
Pope+08 Composite  & 0.14    & 0.19 & $1.3\times10^{5}$ & 2512 & 53.6 & 0.10 & 0.74 & 8.5$\times10^{4}$ & 128 & 660 \\
Brisbin+17         & 0.15    & 1.1 & 1100 & 1899 & 0.591 & $-$0.03 & 4.63 & 22 & 97 & 0.24 \\
\mmpz              & 0.09    & 0.29 & 1600 & 1736 & 0.919 & 0.02 & 0.37 & 237 & 128 & 1.85 \\
\hline\hline
\end{tabular}\\

\begin{minipage}{1.99\columnwidth}
{\bf Table Notes:} $\langle \Delta z/(1+z)\rangle$ captures the
accuracy and $\sigma_{\Delta z/(1+z)}$ captures the precision of each
fitting method.  Negative values of $\langle \Delta z/(1+z)\rangle$
correspond to systematically lower FIR-$z$'s than truth, and positive
values correspond to systematically higher FIR-$z$'s.  Each mock
galaxy test simulates 10$^4$ galaxies, though only some fraction of
those are above the detection thresholds in the simulated bands such
that FIR-$z$'s can be fit.  The number of mock/real galaxies simulated
for each method is reflected in $\nu$, the degrees of freedom, whereby
$\nu=n_{\rm mocks}-1$ or $\nu=n_{\rm real}-1$.  The
last column gives the reduced $\chi^2_{\nu}$; a value close to one
represents appropriate precision on FIR-$z$ uncertainty.  The
spectroscopically-confirmed calibration sample of DSFGs shows the same
trends as the mock galaxy sample though with a much smaller sample;
the \citet{brisbin17a} fit is unique in that many of the
spectroscopically-confirmed galaxies had no valid photometric redshift
due to lack of converged parabolic fit to photometry.
\end{minipage}
\label{tab:1}
\end{table*}

Note that both \citet{ivison16a} and \citet{zavala18c} find a trend
such that $\Delta z/(1+z)$ is systematically lower at higher redshifts
than at lower redshifts.  This `skew' in FIR-$z$ fits is attributed to
the adoption of a single-template SED technique, whereby the
distribution of galaxies sampled have true variance in their SEDs.
While this might be thought to be evidence of redshift evolution in
galaxies' average dust SEDs, the results from Figure~\ref{fig:simple}
instead suggest that the effect is likely driven by the
\lir-\lpeak\ relation.  The effect is especially strong for galaxies
that are bright in the {\it Herschel}-SPIRE bands (as both those
analyzed in \citealt{ivison16a} and \citealt{zavala18c} are).
Galaxies of fixed {\it Herschel} flux densities will be intrinsically
more luminous at higher redshifts, and at higher luminosities they are
more likely to be intrinsically hotter according to the
\lir-\lpeak\ relation. If a single template SED is used, with fixed dust
temperature, then at higher redshifts that template SED is likely
cooler than the galaxies are intrinsically, thus the FIR-$z$ fit is
more likely to peak at lower redshifts than reality.  Such a trend is
only seen in single-temperature SED fits (also visible in
Figure~\ref{fig:simple}), and it is not present in the
\citet{brisbin17a} fits or the \mmpz\ fits.

Figure~\ref{fig:mock} shows two random mock galaxies taken from the
distribution of simulated sources: one near $z\sim2$ and another near
$z\sim6$.  Their photometry vary in SNR from $\sim$10$\sigma$ to
non-detections (whose 1$\sigma$ upper limits are shown) for the
$z\sim2$ mock system and from non-detections to $\sim$35$\sigma$ for
the $z\sim6$ mock system.  The intrinsic SED is traced by square
points and a dotted line, while the mock noise-added photometry, and
resulting best-fit SEDs, shown in gray.  The bottom panels of
Figure~\ref{fig:mock} show the probability density distributions of
each photometric redshift fitting technique analyzed for comparison
with \mmpz.  In the case of the $z\sim2$ source, we see a clear case
of the single-template fits dramatically underestimating the
uncertainties compared to the \citet{brisbin17a} and \mmpz\ estimates,
while the $z\sim6$ case sees a broader distribution in redshift for
each technique, with the \mmpz\ fit providing the most accurate
estimate.

Table~\ref{tab:1} summarizes the precision and accuracy of the fitting
methods tested herein, both for the mock galaxies and for the
spectroscopically-confirmed calibration sample of DSFGs.  The $\langle
\Delta z/(1+z)\rangle$ value gives the aggregate median of the
distribution of $(z_{\rm
    FIR/mm}- z_{\rm real})/(1+z_{\rm real})$ for all sources and
captures the fitting technique's accuracy.  The $\sigma_{\Delta
  z/(1+z)}$ parameter, which is the standard deviation
  in $\Delta z/(1+z)$, captures the breadth of the $\Delta z/(1+z)$
distribution for each fitting method, in other words, the precision of
each fitting technique across large samples.  The
$\chi^2$ parameter is here defined as:
\begin{equation}
\chi^2 = \sum_{i} \frac{(z_{\rm FIR/mm}-z_{\rm real})^2}{(\sigma_{\rm z_{\rm FIR/mm}})^2}
\end{equation}

A reduced chi-squared statistic $\chi^2_\nu$ is quoted by dividing
$\chi^2$ by the number of observations (in this case, independent mock
or real galaxies) less the number of fitted parameters (in this case,
only the redshift); thus, the degrees of freedom $\nu$ is equal to the
number of mock or real galaxies analyzed minus one, ranging between
$\approx$1700--2500 for the mocks.

Note that, assuming the fit is accurate, a reduced chi-squared,
$\chi^{2}_{\nu}$, value far larger than one means that the
uncertainties on the given measurement are under-estimated, while
values significantly less than one imply an over-estimation of
uncertainty.  Our measurements show that single-template fits have
reduced $\chi^2_{\nu}$ values significantly larger than one (of order
$\mathcal{O}(\chi^2_{\nu})\approx$10$^1$--10$^3$), reflective of the
under-estimation of the uncertainties in the FIR-$z$.  The
\citet{brisbin17a} method has $\chi^{2}_{\nu}=0.59$ for the mocks and
$\chi^{2}_{\nu}=0.24$ for the spec-$z$ sample, suggesting an
over-estimation of the uncertainties.  The \mmpz\ method has
$\chi^2_{\nu}=0.92$ for the mocks and $\chi^2_{\nu}=1.85$ for the
spec-$z$ sample.  These values are the closest to one, suggesting the
method's uncertainties are accurately estimated.  Of the two methods
that produce $\chi^2_{\nu}\sim1$ --- the \citeauthor{brisbin17a}
method and \mmpz --- the \mmpz\ technique is more accurate ($\langle
\Delta z/(1+z)\rangle<0.1$) and has better precision ($\sigma_{\Delta
  z/(1+z)}\approx0.3-0.4$ vs. $\sigma_{\Delta z/(1+z)}\approx1-5$).

\section{Conclusions}\label{sec:conclusions}

The \mmpz\ algorithm is introduced as a simple yet reliable technique
for deriving a far-infrared/millimeter photometric redshift for
distant galaxies.  Code is made available to the community to fit a
galaxy's photometry and derive a probability density distribution in
redshift based on the source's photometry and the assumption that the
galaxy will likely lie close to the average redshift-independent
\lir-\lpeak\ relationship.  All available photometric
  constraints can be input, including non-detections.

 The \mmpz\ method contrasts to single galaxy template SED fits for
 redshifts because (1) it accounts for the broad distribution in
 intrinsic dust SEDs that galaxies are known to exhibit, and (2) it
 uses galaxies' measured flux densities directly, and not just FIR/mm
 colors, to infer the most likely redshift solution.  The first point
 ensures that the quoted precision of the FIR-$z$ measurement is not
 under-estimated, as is the case when a single-template SED is used.
 The second point addresses the observed trend in \lir-\lpeak. In
 other words, a galaxy with brighter intrinsic flux densities (but the
 same FIR/mm colors) is more likely to have a higher \lir, and thus a
 shorter \lpeak\ (corresponding to an intrinsically hotter SED); an
 intrinsically hotter SED would mean that the source is more likely to
 sit at higher redshifts than a fainter galaxy with the same observed
 FIR/mm colors, though the uncertainty on both predicted redshift
 distributions would be broad.

Using samples of thousands of mock galaxies and 129
spectroscopically-confirmed DSFGs spanning a wide range of redshifts
and intrinsic dust SEDs, I compare the \mmpz\ technique to a variety
of single-template SED FIR-$z$ fits in the literature, in addition to
the technique outlined in \citet{brisbin17a}.  As suspected, the
single-template SEDs are found to dramatically under-estimate the
uncertainties on FIR photometric redshifts.  Both the
\citet{brisbin17a} and \mmpz\ techniques do a good job of properly
capturing the (low) precision of FIR-$z$ fits (with
  reduced chi-squared $\chi^2_{\nu}=0.6-0.9$), though the
\mmpz\ technique is found to have more accurate and precise
($\chi^2_{\nu}\approx0.9-1.8$, $\langle \Delta z/(1+z)\rangle<0.1$,
$\sigma_{\Delta z/(1+z)}\approx0.3-0.4$) FIR-$z$ estimates across a
wide redshift range.

The \mmpz\ FIR-$z$ fitting technique is most useful for galaxies that
lack constraints at wavelengths outside of the FIR/mm regime.  For
example, it may be of optimal use for large, wide-field surveys from
single-dish millimeter telescopes (e.g. surveys from {\sc TolTEC} on
the LMT, or {\sc NIKA-2} on the IRAM-30\,m) or alternatively, galaxies
characterized with ALMA who fall outside of, or drop out from, deep
optical/near-infrared imaging surveys.  It should be emphasized that
FIR-$z$ fitting in general is a ``last resort'' method of obtaining
redshift constraints on distant galaxies, and it remains a highly
uncertain enterprise.  The core assumption at the root of the
\mmpz\ method is that galaxies fall on a redshift-invariant
\lir-\lpeak\ relationship within some statistical scatter; future
measurements of large samples of dust SEDs for high-$z$ galaxies could
reveal this core assumption to be invalid, but at present, existing
measurements support this assumption out to $z\sim6$.  The intention
behind the introduction of the \mmpz\ algorithm is to provide a
straightforward estimate of FIR/mm photometric redshifts that captures
both the estimated redshift and its uncertainty as best as possible.

\vspace{5mm}

\acknowledgements

CMC thanks the anonymous referee for very helpful and constructive
feedback, and Jorge Zavala and Sinclaire Manning for useful
conversations during the preparation of this manuscript.  CMC also
thanks the National Science Foundation for support through grants
AST-1714528 and AST-1814034, and additionally the University of Texas
at Austin College of Natural Sciences for support.  In addition, CMC
acknowledges support from the Research Corporation for Science
Advancement from a 2019 Cottrell Scholar Award sponsored by IF/THEN,
an initiative of Lyda Hill Philanthropies.

\bibliography{caitlin-bibdesk}

\end{document}